\documentclass[dvips]{article}
\usepackage{supertabular,lscape,epsfig}
\usepackage{amssymb}
\usepackage{amsmath}

\DeclareSymbolFont{ppa}{OT1}{ppl}{m}{it}
\DeclareMathSymbol{\vv}{\mathalpha}{ppa}{'166}

\begin{document}

\newcommand{\dd}{\,{\rm d}}
\newcommand{\ie}{{\it i.e.},\,}
\newcommand{\etal}{{\it et al.\ }}
\newcommand{\eg}{{\it e.g.},\,}
\newcommand{\cf}{{\it cf.\ }}
\newcommand{\vs}{{\it vs.\ }}
\newcommand{\zdot}{\makebox[0pt][l]{.}}
\newcommand{\up}[1]{\ifmmode^{\rm #1}\else$^{\rm #1}$\fi}
\newcommand{\dn}[1]{\ifmmode_{\rm #1}\else$_{\rm #1}$\fi}
\newcommand{\upd}{\up{d}}
\newcommand{\uph}{\up{h}}
\newcommand{\upm}{\up{m}}
\newcommand{\ups}{\up{s}}
\newcommand{\arcd}{\ifmmode^{\circ}\else$^{\circ}$\fi}
\newcommand{\arcm}{\ifmmode{'}\else$'$\fi}
\newcommand{\arcs}{\ifmmode{''}\else$''$\fi}
\newcommand{\MS}{{\rm M}\ifmmode_{\odot}\else$_{\odot}$\fi}
\newcommand{\RS}{{\rm R}\ifmmode_{\odot}\else$_{\odot}$\fi}
\newcommand{\LS}{{\rm L}\ifmmode_{\odot}\else$_{\odot}$\fi}

\newcommand{\Abstract}[2]{{\footnotesize\begin{center}ABSTRACT\end{center}
\vspace{1mm}\par#1\par
\noindent
{~}{\it #2}}}

\newcommand{\TabCap}[2]{\begin{center}\parbox[t]{#1}{\begin{center}
  \small {\spaceskip 2pt plus 1pt minus 1pt T a b l e}
  \refstepcounter{table}\thetable \\[2mm]
  \footnotesize #2 \end{center}}\end{center}}

\newcommand{\TableSep}[2]{\begin{table}[p]\vspace{#1}
\TabCap{#2}\end{table}}

\newcommand{\FigCap}[1]{\footnotesize\par\noindent Fig.\  %
  \refstepcounter{figure}\thefigure. #1\par}

\newcommand{\TableFont}{\footnotesize}
\newcommand{\TableFontIt}{\ttit}
\newcommand{\SetTableFont}[1]{\renewcommand{\TableFont}{#1}}

\newcommand{\MakeTable}[4]{\begin{table}[htb]\TabCap{#2}{#3}
  \begin{center} \TableFont \begin{tabular}{#1} #4 
  \end{tabular}\end{center}\end{table}}

\newcommand{\MakeTableSep}[4]{\begin{table}[p]\TabCap{#2}{#3}
  \begin{center} \TableFont \begin{tabular}{#1} #4 
  \end{tabular}\end{center}\end{table}}

\newenvironment{references}%
{
\footnotesize \frenchspacing
\renewcommand{\thesection}{}
\renewcommand{\in}{{\rm in }}
\renewcommand{\AA}{Astron.\ Astrophys.}
\newcommand{\AAS}{Astron.~Astrophys.~Suppl.~Ser.}
\newcommand{\ApJ}{Astrophys.\ J.}
\newcommand{\ApJS}{Astrophys.\ J.~Suppl.~Ser.}
\newcommand{\ApJL}{Astrophys.\ J.~Letters}
\newcommand{\AJ}{Astron.\ J.}
\newcommand{\IBVS}{IBVS}
\newcommand{\PASP}{P.A.S.P.}
\newcommand{\Acta}{Acta Astron.}
\newcommand{\MNRAS}{MNRAS}
\renewcommand{\and}{{\rm and }}
\section{{\rm REFERENCES}}
\sloppy \hyphenpenalty10000
\begin{list}{}{\leftmargin1cm\listparindent-1cm
\itemindent\listparindent\parsep0pt\itemsep0pt}}%
{\end{list}\vspace{2mm}}

\def\TYLDA{~}
\newlength{\DW}
\settowidth{\DW}{0}
\newcommand{\dw}{\hspace{\DW}}

\newcommand{\refitem}[5]{\item[]{#1} #2%
\def\REFARG{#3}\ifx\REFARG\TYLDA\else, {\it#3}\fi
\def\REFARG{#4}\ifx\REFARG\TYLDA\else, {\bf#4}\fi
\def\REFARG{#5}\ifx\REFARG\TYLDA\else, {#5}\fi.}

\newcommand{\Section}[1]{\section{#1}}
\newcommand{\Subsection}[1]{\subsection{#1}}
\newcommand{\Acknow}[1]{\par\vspace{5mm}{\bf Acknowledgements.} #1}
\pagestyle{myheadings}

\newfont{\bb}{ptmbi8t at 12pt}
\newcommand{\xrule}{\rule{0pt}{2.5ex}}
\newcommand{\xxrule}{\rule[-1.8ex]{0pt}{4.5ex}}
\def\thefootnote{\fnsymbol{footnote}}
\begin{center}
{\Large\bf Metallicity dependence of the Blazhko effect.}
\vskip1cm
Rados\l{}aw~~S~m~o~l~e~c
\vskip3mm
Copernicus Astronomical Center, ul.~Bartycka~18, 00-716~Warszawa, Poland\\
e-mail: smolec@camk.edu.pl
\end{center}

\Abstract{The microlensing surveys, such as OGLE or MACHO, have led to the discovery of thousands of RR~Lyrae stars in the Galactic Bulge and in the Magellanic Clouds, allowing for detailed investigation of these stars, specially the still mysterious Blazhko phenomenon. Higher incidence rate of Blazhko (BL) variables in the more metal-rich Galactic Bulge than in the LMC, suggests that occurance of Blazhko effect correlates with metallicity (Moskalik and Poretti 2003). To investigate this problem, we calibrate the photometric method of determining the metallicity of the RRab star (Kov\'acs and Zsoldos 1995) to the {\it I}-band and apply it to the OGLE Galactic Bulge and LMC data. In both systems, metallicities of non Blazhko and Blazhko variables are close to each other. The LMC Blazhko pulsators prefer slightly lower metallicities. The different metallicities of the Galactic Bulge and the LMC, can't explain the observed incidence rates.

As a by-product of our metallicity estimates, we investigate the luminosity--metallicity relation, finding a steep dependence of the luminosity on [Fe/H].}{stars: horizontal branch -- stars: oscillations -- stars: variables: RR~Lyr -- stars: abundances}
\vspace*{12pt}
\Section{Introduction}
The Blazhko effect, discovered for the first time in 1907 for an RR~Lyrae star RW~Draconis (Blazhko 1907), appears as a cyclic modulation of shape and amplitude of the light curve. About 30\% of all galactic, fundamental mode RR~Lyraes (RRab stars) display this effect (Smith 1995). Despite almost a century of study, we still don't understand this phenomenon. 

The microlensing surveys, such as the Optical Gravitational Lensing Experiment (OGLE; Udalski \etal 1992) or MACHO (Alcock \etal 1996), have led recently to the discovery of thousands of RRab stars in the Galactic Bulge and in the Large Magellanic Cloud (LMC), hundreds of them displaying the Blazhko effect. Detailed analysis of the MACHO LMC data (Alcock \etal 2000, 2003) showed, that Blazhko variables are multiperiodic pulsators with several closely spaced frequencies. The primary peak of the spectrum, $f_{0}$, is accompanied by close secondary frequency (dublet) or by two secondary frequencies located symmetrically on both sides of the primary peak (symmetric triplet). This structure is repeated at the harmonics of the primary frequency. There are also nonequidistant triplets and multiplets present in the spectra, but these are rare cases (Alcock \etal 2003). The separation between frequencies, $\Delta f$, is usually very small, typically $\Delta f<0.1$c/d. Corresponding Blazhko periods, $P_{B}=1/\Delta f$, are of order of days to hundreds of days. Following Alcock \etal (2003), we denote singly periodic, fundamental mode variables as RR0-S, variables with dublets as RR0-BL1 and variables with symmetric triplets as RR0-BL2. The earlier analysis (Alcock \etal 2003) showed, that latter two types overlap, indicating the same underlying phenomenon, so the common notation, RR0-BL, will also be used. 

Statistics on recently analyzed objects is given in Table~1.

{\small
\begin{table}
\begin{center}
\caption{Statistics of RRab stars in the Galactic Bulge and in the LMC. $N_{S}$ and $N_{BL}$ denote the number of RR0-S and RR0-BL stars respectively.}

\vspace{0.4cm}

\begin{tabular}{cccccc}
\hline
Object & Catalogue & $N_{S}$ & $N_{BL}$ & \% & Ref. \\
\hline
Galactic   & OGLE-I & 112 & 35 & 23.4\% & Moskalik, P., Poretti, E. 2003\\
Bulge      & OGLE-II & 1556 & 472 & 23.3\% & Mizerski, T. 2003\\
\hline
LMC & OGLE-II & 4608 & 843 & 15.5\% & Soszy\'nski, I. \etal 2003\\
    & MACHO & 4882 & 751 & 13.3\% & Alcock, C. \etal 2003\\
\hline
\end{tabular}
\end{center}
\end{table}
}

It is clearly visible that Blazhko phenomenon is more frequent in the Galactic Bulge, compared to LMC. Moskalik and Poretti (2003) suggested, that this is due to the difference in metallicity, which is higher in the Galactic Bulge. Verification of this hypothesis, which is the goal of this work, requires an effective method of estimating the metallicity of RRab star. Such a method was proposed by Kov\'acs and Zsoldos (1995) for RR0-S variables. Metallicity is determinated on the basis of the Fourier decomposition parameters of the {\it V}-band light curve. The OGLE data, which is the subject of this analysis, is mostly the Cousins {\it I}-band photometry, so the new calibration of this method is necessary. 

Information about the Blazhko pulsators metallicity may be also interesting for theoretical modeling. Currently considered models incorporate nonradial geometry of pulsation (\eg Nowakowski and Dziembowski 2001, Shibahashi 2000), but none can reproduce all of the observed features. Any information about [Fe/H] may be very useful.

In Section~2 we describe in detail the method of estimating iron abundance of the RRab star, its calibration to the {\it I}-band and application to Blazhko variables. In Section~3 the techniques of light curve analysis both for singly periodic and Blazhko variables are described. In Section~4 we analyze the Fourier parameters of the light curves and metallicities of the RRab variables from the OGLE-I Galactic Bulge sample and the OGLE-II LMC sample. The luminosity--metallicity relation for the LMC RR0-S pulsators is examined. Conclusions are collected in Section~5.

\section{Method}

\subsection{The photometric metallicities}

Following the idea, that shape of the light curve depends on physicall parameters of the star, Kov\'acs and Zsoldos (1995) proposed a new method of estimating metallicity of the RRab star, based on its light curve Fourier parameters. The method was further developed by Jurcsik and Kov\'acs (1996, hereafter JK). They found, that the basic relation between [Fe/H] and Fourier parameters is linear and contains only the period and the Fourier phase, $\phi_{31}$. Their method was calibrated on 84 singly periodic, galactic field RRab stars, with good {\it V} light curves and spectroscopic metallicities. All the [Fe/H] values were transformed to a common metallicity scale introduced by Jurcsik (1995) and based on the recent High Dispersion Spectroscopy (HDS) measurments (hereafter refered as the HDS scale). This scale provides consistent metallicities for both field and cluster RR~Lyrae stars. Jurcsik (1995) also gives a new, general $\Delta S\rightarrow\mathrm{[Fe/H]}$ transformation formula. 

The method was tested by JK for 11 Galactic and LMC globular clusters. Photometric metallicities, averaged for several RR~Lyraes, $\langle\mathrm{[Fe/H]_{fit}}\rangle$, were compared to cluster's spectroscopic metallicity, $\mathrm{[Fe/H]_{obs}}$. It was found, that photometric abundances generally fit well spectroscopic values, but in the low metallicity range,  $\mathrm{[Fe/H]_{obs}}<-2.0$, photometric abundances are higher, on avarege by $\langle\langle\mathrm{[Fe/H]_{fit}}\rangle-\mathrm{[Fe/H]_{obs}}\rangle=0.34\pm 0.04$ (3 clusters). No explanation was given for this discrepancy. In higher metallicity range, $\mathrm{[Fe/H]_{obs}}>-1.8$, the discrepancy still exsist, but it is very small, namely  $\langle\langle\mathrm{[Fe/H]_{fit}}\rangle-\mathrm{[Fe/H]_{obs}}\rangle=0.05\pm 0.03$ (8 clusters). Since the HDS scale is not strictly reproduced by photometric method, we will refer to the photometric metallicities obtained from JK formula as based on the JK photometric scale.

\subsection{Calibration in the {\it I}-band}

As a by-product of the OGLE project, a good quality photometry for thousands of RRab variables has been collected. Since this photometry is mainly in the standard Cousins {\it I}-band, estimating metallicity of the star requires new calibration of the photometric method. Unfortunately there is much less good photometry for field RR~Lyrae stars in the {\it I}-band, than in the {\it V}, so the calibration is based only on 28 singly periodic, fundamental mode pulsators\footnote{Comparing the Clementini \etal (1990) {\it V} photometry, with Lub (1977), originaly Walraven photometry, JK claimed, that V Ind shows the Blazhko effect. However extensive Clementini \etal (1990) photometry alone, shows no sign of Blazhko effect, so following these authors, we treat V Ind as a single mode pulsator.}. The HDS metallicity scale is adopted here.

For 21 field RRab stars good {\it I} light curves are available. For three of them, DX Del, V445 Oph and VY Ser, photometry is in the Johnson's {\it I}-band. Transformation to the standard Cousins {\it I} is done, using Bessel (1983) formulae. Except V Ind, all the stars are also in JK calibrating sample, so their metallicities are adopted. Metallicity for V Ind was obtained by Layden (1994) and transformation to HDS scale is done according to formulae given in JK (their equation 2).  

Photometry for another six variables comes from $\omega$ Centauri (Ka\l{}u\.zny \etal 1997), which is the globular cluster with high metallicity spread. Photometry is given in the instrumental system, but according to Ka\l{}u\.zny \etal (1996) it reproduces the Cousins {\it I}-band fairly well, and transformation is not necessary. Metallicities of these variables were obtained using Ca{\it by} photometry technique (Rey \etal 2000), calibrated on Layden (1994) stars, so the transformation to the HDS scale is done just as above, using JK transformation.

The last variable comes from NGC 6362 (Mazur \etal 1999), with metallicity obtained by Costar and Smith (1988) and transformed to the HDS scale by Jurcsik (1995).

All objects with light curve reference and [Fe/H] value are presented in Table~2.
{\small
\begin{table}
\begin{center}
\caption{The calibrating data set. Globular cluster star Id's are consistent with Sawyer-Hogg (Hogg 1973). Light curve references: (1) Liu, T., Janes, K.A. 1989; (2) Skillen, I. \etal 1993; (3) Hansen, L., Petersen, J.O. 1991; (4) Barnes, T.G. \etal 1988; (5) Clementini, G. \etal 1990; (6) Fernley, J.A. \etal 1990; (7) Cacciari, C. \etal 1987; (8) Mazur, B. \etal 1999; (9) Ka\l{}u\.zny, J. \etal 1997.}

\vspace{0.4cm}

\begin{tabular}{cccccc}
\hline
Star  & Ref. & [Fe/H] & Star & Ref. & [Fe/H]\\
\hline
SW And  & 1 & $-0.06$ & AV Peg   & 1 & $+0.08$\\
WY Ant  & 2 & $-1.39$ & AR Per   & 1 & $-0.14$\\
V Cae   & 3 & $-1.71$ & BB Pup   & 2 & $-0.35$\\
RR Cet  & 1 & $-1.29$ & VY Ser   & 6 & $-1.58$\\
W Crt   & 2 & $-0.45$ & V440 Sgr & 7 & $-1.21$\\
DX Del  & 4 & $-0.32$ & W Tuc    & 5 & $-1.37$\\
SU Dra  & 1 & $-1.56$ & TU UMa   & 1 & $-1.15$\\
RX Eri  & 1 & $-1.07$ & NGC 6362 v19 & 8 & $-0.83$\\
V Ind   & 5 & $-1.24$ & $\omega$ Cen v40  & 9 & $-1.33$\\
RR Leo  & 1 & $-1.30$ & $\omega$ Cen v44  & 9 & $-1.14$\\
SS Leo  & 6 & $-1.56$ & $\omega$ Cen v51  & 9 & $-1.37$\\
TT Lyn  & 1 & $-1.50$ & $\omega$ Cen v62  & 9 & $-1.35$\\
RV Oct  & 2 & $-1.08$ & $\omega$ Cen v102 & 9 & $-1.56$\\
V445 Oph & 4 & $+0.01$ & $\omega$ Cen v115 & 9 & $-1.59$\\
\hline
\end{tabular}
\end{center}
\end{table}
}

Each light curve is fitted with the Fourier series of the form:
\begin{equation}
I(t)=A_{0}+\sum_{k=1}^{N}A_{k}\sin(2\pi kft+\phi_{k})
\label{series}
\end{equation}
The order of the decomposition, $N$, is chosen to satisfy the condition: $A_k/\sigma_{A_k}>4$, for all the amplitudes. $\sigma_{A_k}$ stands for the standard error of the amplitude, $A_k$. The pulsation frequency, $f$, is adjusted to obtain the minimal value of the fit's dispersion, $\sigma$. In the case of WY Ant, optimal period is strongly dependent on the decomposition order, so the Kholopov (1995) period is adopted. Periods of $\omega$ Centauri variables are derived from the more extensive {\it V} photometry. The TU UMa period is obtained by phasing the data from two sources (Liu and Janes 1989, Barnes \etal 1988). In all cases, points deviating from the fit by more than $4\sigma$ are rejected.

Linear combinations of the Fourier parameters, up to the fifth order, namely the period, $P$, the amplitudes, $A_{1}-A_{5}$, and the Fourier phases, $\phi_{21}-\phi_{51}$ ($\phi_{k1}=\phi_{k}-k\phi_{1}$), are tested to fit the observed metallicities, $\mathrm{[Fe/H]_{obs}}$. The best two-parameter formula contains, like in JK, the period and the $\phi_{31}$ phase:
\begin{equation}
{\setlength\arraycolsep{3pt}
\begin{array}{lllll}
\mathrm{[Fe/H]}= &-3.142    &-4.902P & +0.824\phi_{31} &\\ 
                 &\pm 0.646 &\pm 0.375 &\pm 0.104 &\ \ \ \sigma=0.18
\end{array}}
\label{calI2}
\end{equation}
$\sigma$ denotes dispersion of the fit. Abundances determined from above equation will be denoted by $\mathrm{[Fe/H]}_{2}$. The three-parameter formulae have smaller dispersions. The best one, contains additionaly amplitude $A_2$:
\begin{equation}
{\setlength\arraycolsep{3pt}
\begin{array}{llllll}
\mathrm{[Fe/H]}= &-6.125    &-4.795P   & +1.181\phi_{31} & +7.876A_{2}&\\
                 &\pm 0.832 &\pm 0.285 &\pm 0.113        &\pm 1.776&\ \ \ \sigma=0.14
\end{array}}
\label{calI3}
\end{equation}
Abundances determined from above equation will be denoted by $\mathrm{[Fe/H]}_{3}$. Formulae using four or more parameters are much worse. 

Observed \vs fitted abundances for the calibrating sample are presented on Fig.~1. 

\subsection{Application to Blazhko variables}

Light curve shape of the singly periodic RRab variables, used in the calibration, is time-independent. In the case of the Blazhko variables, shape of the light curve varies from season to season. Light curve phased with the fundamental period is characterized by a large scatter, specially in the light maximum (Fig.~2a). Decomposition of such a curve is useless. Jurcsik, Benk\H o and Szeidl (2002) showed, that also seasonal curves, at different phases of the Blazhko cycle, don't fit the typical singly periodic curve. However, we can extract the fundamental mode light curve, by prewithening the data with the additional close frequencies (Fig.~2b). We assume, that Fourier parameters of this light curve may be used to estimate the Blazhko variable metallicity. 

To prove this assumption, one should have some Blazhko variables with spectroscopic metallicities, and extensive photometry, allowing for detailed Fourier analysis and filtering out additional close components, in order to estimate photometric metallicity for comparison. Such a {\it I}-band photometry is available for the Galactic Bulge and the Magellanic Clouds OGLE samples, but there is only one individual metallicity value for Blazhko variable of the LMC, that is also in our sample, namely SC\_21 16249 (OGLE Id)\footnote{We mention that also Borissova \etal (2004) gives individual metallicity values for some LMC RRab stars, six of them being Blazhko variables (Alcock \etal 2003). However they mix the spectroscopic values with photometric metallicities (obtained from {\it V} light curves) and hence we can't use their values, since we do not know, which ones are spectroscopic.}. Spectroscopic metallicity value comes from Gratton \etal (2004), and is transformed to HDS scale as described in Section~4.2, leading to value $-1.63\pm 0.11$. Estimated, photometric metallicity (using Eq. 2, \cf Sec.~4.2) is $-1.35\pm 0.18$, so the difference between these values is $-0.28\pm 0.21$ - quite large, but still acceptable (1.3$\sigma$).

Of course one star is unsufficient to test the method, however, for some galactic field, Blazhko pulsators, extensive and homogeneous {\it V} photometry suitable for Fourier analysis exists, and can be used to test the method in case of the {\it V} band. Unfortunately some of them (V442 Her, Schmidt and Lee 2000; GV And, Lee \etal 2002) don't have spectroscopic [Fe/H], or value is very uncertain (AH Cam, Smith \etal 1994). In the case of XZ Dra, and RS Boo, available homogeneus sets of {\it V} photoelectric measurments (Szeidl \etal 2001, Kany\'o 1986) were made for O--C analysis, so only the rising branch and maximum of light are well covered. We are left with only five stars, for which we are able to test the method, namely: DR And (Lee and Schmidt 2001), XZ Cyg (LaCluyz\'e \etal 2004), AR Her (Smith \etal 1999), RR Lyr (Smith \etal 2003) and RV UMa (Kov\'acs 1995). For RV UMa detailed Fourier analysis of {\it V} light curve was already done by Kov\'acs (1995), so we use their extracted Fourier parameters of the fundamental mode light curve (their Table 3) to calculate the photometric metallicity. In case of DR And, XZ Cyg, AR Her and RR Lyr we perform the same analysis as described in Section~3 (and in greater detail in Appendix) to extract Fourier paramameters of the fundamental mode light curve. Calculated photometric metallicities (JK's Eq.~3) are presented in column 3 of Table 3. Their errors are assumed to be 0.15 dex (JK, Jurcsik 1998 and references therein).

In case of RR Lyr, XZ Cyg and RV UMa, spectroscopic metallicities are weighted averages of the Layden (1994) and Suntzeff (1994) values, transformed to the HDS scale as described in JK. We used the same weightening scheme as JK, to set the metallicities on the same scale, as the {\it V} photometric method was calibrated. Metallicity for DR And comes from Layden (1994) and is transformed as in JK. In case of AR Her spectroscopic [Fe/H] is derived from Preston (1959) $\Delta S$, using transformation given by Jurcsik (1995). Spectroscopic [Fe/H] values are collected in column 2 of Table 3 together with errors of the weighted averages. In case of AR Her the error coresponds to the uncertainity of one unit in $\Delta S$. Difference between photometric and spectroscopic values is presented in column 4. 

In case of DR And, XZ Cyg, AR Her and RR Lyr there is a nice agreement between estimated and observed [Fe/H]. Only in case of RV UMa the difference is larger, but acceptable (1.2$\sigma$).

We conclude that the described method may be used to obtain reliable values of Blazhko variable metallicity in case of the {\it V} light curve and assume that it also holds for the {\it I} band, which is used through this paper.
{\small
\begin{table}
\begin{center}
\caption{Spectroscopic \vs {\it V}-photometric  metallicities for Blazhko variables.}

\vspace{0.4cm}

\begin{tabular}{cccc}
\hline
Star & $\mathrm{[Fe/H]_{spec}}$  & $\mathrm{[Fe/H]_{phot}}$ & $\Delta\mathrm{[Fe/H]}$ \\ 
\hline
DR And & $-1.22\pm 0.07$ & $-1.14\pm 0.15$ & $-0.08\pm 0.17$\\
XZ Cyg & $-1.18\pm 0.05$ & $-1.07\pm 0.15$ & $-0.11\pm 0.16$\\
AR Her & $-1.11\pm 0.19$ & $-1.14\pm 0.15$ & $\ \ 0.03\pm 0.24$\\
RR Lyr & $-1.12\pm 0.01$ & $-1.10\pm 0.15$ & $-0.02\pm 0.15$\\
RV UMa & $-0.90\pm 0.11$ & $-1.13\pm 0.15$ & $\ \ 0.23\pm 0.19$\\
\hline
\end{tabular}
\end{center}
\end{table}
}

\section{The data and techniques of light curve analysis}

{\bf Singly periodic variables.} The OGLE-I catalogue (Udalski \etal 1994, 1995a, 1995b, 1996, 1997), contains 112 RR0-S stars. The OGLE-II LMC sample (Soszy\'nski \etal 2003), is almost 40 times more numerous, and contains 4607 singly periodic, fundamental mode pulsators. Procedure of the analysis is the same for both samples. Light curves are fitted with the Fourier series, with period adjustment and bad data points rejection, just as described in Section~2.2. Since the data covers multiple seasons, possible instrumental drifts are modeled by a sine term with period of 50,000 days added to Eq.~(1). 

From the Galactic Bulge sample we exclude MM7\_V63, which is probably a member of the Sagittarius Dwarf Galaxy (Olech 1997). In case of the LMC sample, about 200 stars have noticably higher apparent luminosities, than the others. Those which are half a magnitude brighter than the sample average for a given period, are treated as blends and excluded from the sample. Only a good quality curves, that are decomposed to the third or higher order are used in later analysis. This is required by the method of estimating metallicity of the star. 90 RR0-S variables from the Galactic Bulge and 3990 RR0-S from the LMC passed the described criterions.

{\bf Blazhko variables.} In order to estimate the Blazhko variable abundance, the Fourier parameters of the fundamental mode must be extracted. This is done by fitting the light curve with the Fourier series, containing all the frequencies, also that corresponding to the additional close components. Then, the parameters corresponding to the primary frequency and its harmonics, are chosen to the further analysis. 

The Galactic Bulge sample was already studied in detail by Moskalik and Poretti (2003). They gave accurate frequencies and all the Fourier terms present in the spectra for 35 Blazhko variables present in the OGLE-I sample. We just use their Fourier series form and choose the proper parameters to the further analysis. Five stars with more complicated features (not being simple BL1 or BL2 type, but for example, nonequidistant triplet, variable frequency values), are also analyzed.  

The case of the LMC data is more complicated. Soszy\'nski \etal (2003) published only the list of multiperiodic variables, with primary frequency, and one additional close frequency of the highest amplitude, not distinguishing between RR0-BL1 and RR0-BL2 types. Extraction of the primary mode parameters is done by an automatic procedure, that searches only for BL1 or BL2 types. Details of the procedure are given in the Appendix together with a test of the algorithm based on known results from the Galactic Bulge Blazhko pulsators.

Not counting the blends (excluded on the basis of the same criterion as above), 702 LMC stars passed the algorithm criterions, 58\% being of type BL1, and 42\% of type BL2. Corresponding numbers in the MACHO sample (Alcock \etal 2003) are 55\% (BL1) and 45\% (BL2), so the results are consistent.  

\section{Results}

\subsection{The Galactic Bulge pulsators}

The singly periodic RR~Lyrae pulsators from the OGLE-I sample, were already analyzed by Poretti (2001). Moskalik and Poretti (2003) considered the Blazhko pulsators. 

For all the stars, abundances are estimated using both two and three-parameter formulae. Metallicity distributions are presented on Fig.~3. In the case of singly periodic pulsators both formulae  (Eq. 2, 3) give consistent results. Averaged metallicities are $\langle\mathrm{[Fe/H]_2}\rangle=-1.04\pm 0.03$ and $\langle\mathrm{[Fe/H]_3}\rangle=-1.05\pm 0.03$.

Estimated abundances are contained approximately in a range $(-1.4,-0.8)$ for the RR0-S stars. Since the accuracy of the two-parameter formula is only $\sigma=0.18$, it is necessary to check, whether the spread of estimated [Fe/H] is physically significant. We do this by constructing the period--amplitude diagram (the Bailey diagram). It is a well known fact (Smith 1995), that for a given period, the RRab stars of lower metallicity have higher amplitudes. As we can see on Fig.~4, using metallicities obtained with our formulae, we reproduce this relation satisfactorily. Dispersions of metallicities (Fig.~3), larger than the error of the method, indicate intrinsic metallicity spread, of the order of $\sim 0.16$ (using two-parameter formula). 

Both of our [Fe/H] estimating formulae contains Fourier phase, $\phi_{31}$. It is interesting to see where do stars of different [Fe/H], fall in the period--phase diagram. Plots of $\phi_{31}$ and $\phi_{41}$ \vs period for RR0-S stars, are presented on Fig.~5. Of course in case of two-parameter formula (left panel of Fig.~5) and $\phi_{31}$ plot, lines of equal [Fe/H] are straight. The main sequence is formed by stars of [Fe/H] close to the sample average. The area above and below this sequence, is occupied by stars of higher and lower [Fe/H] respectively. This picture remains true for higher order phases as can be seen on $\phi_{41}$ plot. Also addition of $A_2$ amplitude to the [Fe/H] estimating formula (right panel of Fig.~5), doesn't change this picture significantly. Connection between location of the star on the period--phase diagram and its metallicity, for the same Galactic Bulge sample, was already noticed by Poretti (2001). The author distinguished three discrete {\it tails}, which he connected with different [Fe/H]. On the base of our study, we may state, that better description incorporates areas of different [Fe/H], that are parallel to the sequence formed by most of the system stars, with continuous transition between them.

Now, let us compare some features of the non-Blazhko and Blazhko variables light curves. On Figs.~6 and 7, the Fourier amplitude of the fundamental mode, $A_1$, and higher order amplitudes, $A_2-A_5$, and phases, $\phi_{21}-\phi_{51}$, are plotted against period. The pulsation periods range from $0.3$ to $0.9$ day, being shorter than 0.7 days for Blazhko pulsators. On the plots of Fourier phases and $A_{1}$ \vs period, there is no clear distinction between RR0-S and RR0-BL variables. However on the plots of $A_2-A_5$, it is visible, that BL variables prefer lower values of amplitudes. The effect is clearly visible, if average values, for different period bins, are calculated. Results are presented in Table~4. For all period bins, all amplitudes are smaller for BL pulsators. The effect is weakest for $A_1$. The plot of amplitudes $A_2-A_5$, \vs the amplitude, $A_1$, is presented on Fig.~8. For low $A_1$ values, Blazhko pulsators and singly periodic stars follow the same sequence, but for $A_1>0.2$, Blazhko pulsators prefer lower amplitudes of harmonics.

{\small
\begin{table}
\begin{center}
\caption{Averaged Fourier amplitudes and phases in three period bins, both for singly periodic and Blazhko variables from the Galactic Bulge. The error of the mean is given in a parenthesis.}

\vspace{0.4cm}

\begin{tabular}{ccccccc}
\hline
           & \multicolumn{2}{c}{$P\le 0.5$} & \multicolumn{2}{c}{$P\in(0.5,0.6\rangle$} &\multicolumn{2}{c}{$P>0.6$}\\
           & RR0-S     & RR0-BL  & RR0-S       & RR0-BL & RR0-S  & RR0-BL\\
\hline
$A_{1}$    & 0.251{\scriptsize (5)}& 0.227{\scriptsize (5)} & 0.203{\scriptsize (6)} & 0.20{\scriptsize (1)} & 0.162{\scriptsize (9)}& 0.13{\scriptsize (2)}\\
$A_{2}$    & 0.128{\scriptsize (3)}& 0.097{\scriptsize (6)} & 0.106{\scriptsize (4)} & 0.093{\scriptsize (7)} & 0.076{\scriptsize (6)}& 0.06{\scriptsize (1)}\\
$A_{3}$    & 0.091{\scriptsize (3)}& 0.051{\scriptsize (6)} & 0.070{\scriptsize (3)} & 0.057{\scriptsize (5)} & 0.045{\scriptsize (5)}& 0.029{\scriptsize (6)}\\$A_{4}$    & 0.059{\scriptsize (2)}& 0.035{\scriptsize (4)} & 0.044{\scriptsize (2)} & 0.033{\scriptsize (4)} & 0.026{\scriptsize (3)}& 0.021{\scriptsize (4)}\\$A_{5}$    & 0.040{\scriptsize (2)}& 0.024{\scriptsize (4)} & 0.028{\scriptsize (2)} & 0.025{\scriptsize (2)} & 0.018{\scriptsize (2)}& 0.013{\scriptsize (-)}\\\hline
$\phi_{21}$ & 2.57{\scriptsize (2)} & 2.59{\scriptsize (3)} & 2.78{\scriptsize (3)} & 2.84{\scriptsize (5)} & 3.12{\scriptsize (3)}& 3.14{\scriptsize (8)}\\
$\phi_{31}$ & 5.37{\scriptsize (5)} & 5.25{\scriptsize (9)} & 5.83{\scriptsize (4)} & 5.8{\scriptsize (1)}  & 6.39{\scriptsize (7)}& 6.4{\scriptsize (1)} \\
$\phi_{41}$ & 1.92{\scriptsize (7)} & 2.3{\scriptsize (6)}  & 2.58{\scriptsize (7)} & 2.9{\scriptsize (4)}  & 3.4{\scriptsize (1)} & 3.4{\scriptsize (3)} \\
$\phi_{51}$ & 4.73{\scriptsize (9)} & 4.3{\scriptsize (4)}  & 5.59{\scriptsize (9)} & 5.0{\scriptsize (1)}  & 6.5{\scriptsize (2)} & 6.7{\scriptsize (-)} \\
\hline
\end{tabular}
\end{center}
\end{table}
}

Noticed relations among Fourier amplitudes are reflected in metallicity distributions for Blazhko pulsators (Fig.~3). The two-parameter formula gives the similar shape of the distribution as for the singly periodic pulsators, with average $\langle\mathrm{[Fe/H]_2}\rangle=-1.01\pm 0.05$. The three-parameter formula gives significantly lower value, $\langle\mathrm{[Fe/H]_3}\rangle=-1.21\pm 0.07$. This result is not unexpected. Since the three-parameter formula contains $A_2$ amplitude (Eq.~3), which is on average lower for BL stars (\cf Figs.~7 and 8, Table~4), the calculated abundances are smaller. 

We have to deal with a question, which formula, two or three-parameter, gives proper results. It is a question, whether the lower values of $A_2$ amplitudes for BL pulsators, are caused by their lower [Fe/H], or are the consequence of the Blazhko phenomenon, leading to the change of the fundamental mode shape. There are two arguments in favour of second explanation. First, the two-parameter [Fe/H] estimator (which has the higher dispersion only), doesn't lead to the lower abundance of BL stars. Second, as it is visible on Fig.~8, the interrelations between Fourier amplitudes of BL pulsators are disturbed. Resulting conclusion is, that lower values of amplitude $A_2$, are not directly connected with metallicity, but are rather a dynamic effect. In the case of BL pulsators we should use only the two-parameter estimator of [Fe/H]. We have already shown, that using the two-parameter ($P,\ \phi_{31}$) estimator in case of the {\it V} light curves, leads to metallicities consistent with spectroscopic values.  We will return to the differences between our [Fe/H] estimators, while analyzing the LMC sample.

No matter which formula is used, the Blazhko variables do not seem to prefer higher metallicities as suggested by Moskalik and Poretti (2003).      
 
\subsection{The LMC pulsators}

Metallicity distributions of the LMC RRab stars are presented on Fig.~9. Unexpected result is visible for RR0-S stars metallicity distributions. In the case of the Galactic Bulge, both two and three-parameter [Fe/H] estimators lead to consistent results. In the LMC case, the distribution based on a three-parameter [Fe/H] estimator is broader and has a lower average value, $\langle\mathrm{[Fe/H]_3}\rangle=-1.387\pm 0.005$,  while $\langle\mathrm{[Fe/H]_2}\rangle=-1.218\pm 0.004$. The three-parameter [Fe/H] estimator leads to systematically lower metallicities, than the two-parameter one, with the difference increasing with decreasing [Fe/H].

Empirically, we may check which [Fe/H] estimator, two or three-parameter, gives proper results, by comparing with spectroscopic [Fe/H]. According to Clementini \etal (2003), average spectroscopic metallicity of 101 LMC RR~Lyrae stars is $-1.48\pm 0.03$ on the Harris (1996) metallicity scale. Individual [Fe/H] values for these stars are given in Gratton \etal (2004) who compared also spectroscopic [Fe/H] with photometric [Fe/H] for 19 RRab stars. Photometric metallicities were derived from {\it V} light curve and JK formula. It was found that the average difference between the JK photometric scale and the Harris scale is $0.28\pm0.05$. In the considered metallicity range ($\mathrm{[Fe/H]}>-1.8$) JK photometric metallicities are on average higher by $0.05\pm 0.03$ than the HDS metallicities (\cf Section~2.1). Taking this two effects into account, we may state, that the HDS scale is more metal-rich than the Harris scale by $0.23\pm 0.06$. Converting metallicity given by Clementini \etal (2003), from Harris scale to HDS scale (used in this paper), we find $-1.25\pm 0.07$. Comparing this value with average [Fe/H] derived from our formulae, we conclude that the value obtained with two-parameter formula is in very good agreement with spectroscopic [Fe/H], which is not the case for three-parameter formula. We conclude that the two-parameter formula is a better metallicity estimator for the LMC RR0-S stars.

The difference between metallicities obtained from formulae with and without $A_2$ for the RR0-S LMC stars, not noticed in the case of Galactic Bulge, seems to be connected with physical properties of the stars of these systems. The comparison of pulsational properties of stars, being members of our calibrating sample, the Galactic Bulge and the LMC, is done, by construction of $A_2$ \vs $A_1$ plot -- Fig.~10. For clarity, on the left panel we compare the calibrating and the Galactic Bulge samples, while on the right panel the Galactic Bulge and the LMC samples. Stars of the calibrating sample and the Galactic Bulge follow the same progression, while progressions of the LMC stars and the Galactic Bulge stars are inclined to each other. The effect is small, but visible, and is most pronounced for $A_2(A_1)$ relation. For this reason we should not use the three-parameter [Fe/H] estimator to the LMC stars. Only the formula without amplitude $A_2$, will be used in further considerations.

It appears that photometric method, is not at all universal, but slightly depends on the considered stellar system. This was not previously noticed, simply because the JK formula, doesn't  include amplitudes, and it seems, that they are not necessary, when one uses the {\it V} light curves. On the base of this study, we may state, that amplitudes are probably important in the case of {\it I} light curves, however a more detailed calibration, based on a larger, more homogeneous sample is required. Special care should be taken, while applying the method to different stellar systems. 

Fourier amplitudes and phases \vs period, for singly periodic and Blazhko variables, are presented on Figs.~11 and 12. For clarity only one third of the stars are plotted. Similarly as for the Galactic Bulge stars, amplitudes of RR0-BL stars are smaller than for the RR0-S stars. Average Fourier parameters are presented in Table~5. The phases of both RR0-BL and RR0-S pulsators (Fig.~12), seem to follow the same sequence, however it is clear from Table~5, that indeed phases are also smaller for Blazhko pulsators. The smallest difference we note for the low order parameters, $A_2$ and $\phi_{21}$. Similar results were found by Alcock \etal (2003), for their MACHO LMC sample.  

{\small
\begin{table}
\begin{center}
\caption{Averaged Fourier amplitudes and phases in three period bins, both for singly periodic and Blazhko variables from the LMC. The error of the mean is given in a parenthesis}

\vspace{0.4cm}

\begin{tabular}{ccccccc}
\hline
           & \multicolumn{2}{c}{$P\le 0.5$} & \multicolumn{2}{c}{$P\in(0.5,0.6\rangle$} &\multicolumn{2}{c}{$P>0.6$} \\
           & RR0-S     & RR0-BL  & RR0-S       & RR0-BL & RR0-S  & RR0-BL\\
\hline
$A_{1}$    & 0.237{\scriptsize (2)}& 0.227{\scriptsize (3)} & 0.196{\scriptsize (1)} & 0.190{\scriptsize (2)} & 0.153{\scriptsize (1)}& 0.147{\scriptsize (4)}\\$A_{2}$    & 0.114{\scriptsize (1)}& 0.097{\scriptsize (2)} & 0.094{\scriptsize (1)} & 0.081{\scriptsize (1)} & 0.071{\scriptsize (1)}& 0.064{\scriptsize (2)}\\$A_{3}$    & 0.081{\scriptsize (1)}& 0.056{\scriptsize (2)} & 0.0670{\scriptsize (4)} & 0.051{\scriptsize (1)} & 0.045{\scriptsize (1)}& 0.040{\scriptsize (1)}\\
$A_{4}$    & 0.056{\scriptsize (1)}& 0.037{\scriptsize (1)} & 0.0452{\scriptsize (3)} & 0.034{\scriptsize (1)} & 0.0314{\scriptsize (4)}& 0.028{\scriptsize (1)}\\
$A_{5}$    & 0.0410{\scriptsize (4)}& 0.031{\scriptsize (2)} & 0.0323{\scriptsize (2)} & 0.027{\scriptsize (1)} & 0.0232{\scriptsize (4)}& 0.021{\scriptsize (1)}\\\hline
$\phi_{21}$ & 2.541{\scriptsize (6)} & 2.55{\scriptsize (1)} & 2.698{\scriptsize (3)} & 2.689{\scriptsize (7)} & 2.975{\scriptsize (5)}& 2.93{\scriptsize (1)}\\$\phi_{31}$ & 5.27{\scriptsize (1)} & 5.18{\scriptsize (3)} & 5.613{\scriptsize (7)} & 5.50{\scriptsize (1)}  & 6.15{\scriptsize (1)}& 6.05{\scriptsize (3)} \\
$\phi_{41}$ & 1.76{\scriptsize (1)} & 1.66{\scriptsize (5)}  & 2.25{\scriptsize (1)} & 2.09{\scriptsize (3)}  & 2.98{\scriptsize (2)} & 2.74{\scriptsize (6)} \\$\phi_{51}$ & 4.50{\scriptsize (2)} & 4.50{\scriptsize (6)}  & 5.13{\scriptsize (1)} & 4.99{\scriptsize (4)}  & 5.82{\scriptsize (3)} & 5.61{\scriptsize (9)} \\\hline
\end{tabular}
\end{center}
\end{table}
}

It is also worth noticing, that Blazhko pulsators prefer shorter periods. Only 7 Blazhko stars have period longer than 0.7 days. As can be seen from period distribution, presented on Fig.~13, Blazhko variables with periods shorter than 0.6 days, constitute 17\% of all RRab stars. For longer periods, there is a significant decrease of Blazhko effect incidence rate. Similar tendency is seen in the Galactic Bulge.  

Using the two-parameter [Fe/H] estimator, we can now compare the metallicity distributions of RR0-S and RR0-BL stars (left panel of Fig.~9). The average [Fe/H] values are $-1.218\pm 0.004$ and $-1.28\pm 0.01$ respectively, so they are similar, with average being lower for BL variables. The metallicity distribution for RR0-S stars is asymmetric, with a metal-rich {\it tail}, while the distribution for RR0-BL stars is approximately symmetric.  

Lower average metallicity for RR0-BL stars and a metal-rich tail for RR0-S stars, suggests, that Blazhko variables may prefer lower metallicities, opposite to Moskalik and Poretti (2003) suggestion. To verify this hypothesis we divide all the stars into three groups with different metallicities, and in each group we calculate the incidence rate for RR0-BL stars. Results are collected in Tables~6A--C. To avoid possible dependence of result, on the way groups are defined, division is made in three different manners: Table~6A -- groups with approximately equal number of RR0-BL stars, Table~6B -- groups with approximately equal number of RR0-S stars and Table~6C -- central group concentrated on the average metallicity of the RR0-S stars, with the width 0.4, and remaining two groups corresponding to the wings of the metallicity distributions.

Regardless of the division scheme, the RR0-BL stars are most frequent ($19-20\%$) in the lowest metallicity group, with [Fe/H] lower than approximately $-1.4$. In the remaining two groups of higher metallicity ($\mathrm{[Fe/H]}>\approx -1.4$) incidence rates are approximately the same ($12-14\%$).
       
Such an effect is not visible in the case of the Galactic Bulge stars. Here, due to the less numerous sample, we divide the stars into two groups, with metallicities greater and lower than the average [Fe/H] for RR0-S stars. Obtained incidence rates of Blazhko variables are similar: $26\pm 6\%$ in the low metallicity group and $30\pm 6\%$ in high metallicity group. Taking into account the metallicity spread in the Galactic Bulge, $\sim (-1.4,\ -0.8)$, this result is consistent with the LMC result, in the sense that for [Fe/H] higher than $\approx -1.4$, the incidence rate for BL stars is approximately constant.      

We have also checked the relations between [Fe/H] and dynamical parameters characterizing the Blazhko phenomenon, namely the period of the Blazhko cycle, $P_B$ , and the amplitude of the highest additional frequency componenet, $A_B$. No clear dependence was found.

\subsection{Luminosity -- metallicity relation for the LMC RR0-S pulsators}

Having determined metallicity for thousands of RR Lyrae stars we are able to derive the luminosity-metallicity relation with very high statistical accuracy. This relation is often assumed to be linear of the following form: $M_{V}=\alpha\mathrm{[Fe/H]}+\beta$, however there is no agreement on the exact values of the $\alpha$ and $\beta$ parameters. Derived values of the slope fall in a range $0.2-0.3$ (see Olech \etal 2003 and references therein), while zero point estimates fall in a range $0.5-0.7$ at $\mathrm{[Fe/H]}=-1.5$ with the average value from ten different methods equal $0.59\pm 0.03$ (see Cacciari and Clementini 2003 for review).
 
The $V_0-\mathrm{[Fe/H]}$ relation for singly periodic LMC RRab stars is presented on Fig.~14. $V_0$ values are extinction free magnitudes directly taken from OGLE catalog (Soszy\'nski \etal 2003). We have omitted stars from OGLE fields SC17--SC20. They have on average lower magnitudes and form a more distant part of the warped LMC bar (Subramaniam 2003). For stars being overlaped on different fields, independent $V_0$ measurments are averaged. Linear fit to the data (3397 RRab stars) leads to relation:

\begin{equation}
{\setlength\arraycolsep{3pt}
\begin{array}{lll}
V_0= &0.30\mathrm{[Fe/H]}  &+19.33  \\ 
     &\pm 0.02              &\pm 0.02 
\end{array}}
\end{equation}
No evidence of nonlinearity is found. Obtained slope is rather steep, and agrees well with Olech \etal (2003) value ($0.26\pm 0.08$), but is higher than values given by Clementini \etal (2003) ($0.21\pm 0.05$) or Fernley \etal (1998) ($0.18\pm 0.03$).

To compare our estimation of zero point with those collected in Cacciari and Clementini (2003), we note, that their value of $\mathrm{[Fe/H]}=-1.5$ corresponds to $\Delta S=7.03$ (Fernley and Barnes 1997) which corresponds to [Fe/H]=-1.31 on HDS scale (Jurcsik 1995). Our relaton leads to $V_0=18.939\pm 0.005$ at $\mathrm{[Fe/H]}=-1.31$. The absolute magnitude value depends on the assumed distance modulus to the LMC, $\mu_{LMC}$, but there is a large spread in the estimated values (see Clementini \etal 2003, specially their Fig.~8). Adopting the preferred value used by HST Key Project, $\mu_{LMC}=18.5\pm 0.1$, (Freedman \etal 2001) we obtain $0.44\pm 0.10$ mag for the zero point in absolute magnitude (at $\mathrm{[Fe/H]}=-1.31$), brighter than the value given by Cacciari and Clementini (2003) ($0.59\pm 0.03$). The difference may be due to overestimated OGLE reddenings given in Udalski \etal (1999) and adopted in Soszy\'nski \etal (2003) catalog, as noticed in the literature (\eg Clementini \etal 2003). The average OGLE reddening is: $E(B-V)=0.143$, to be compared with values adopted by HST Key Project (Freedman \etal 2001) and Fouqu\'e \etal (2003), $E(B-V)=0.1$, value given by Laney and Stobie (1994), $E(B-V)=0.07$, or Subramaniam (2005), $E(B-V)=0.06$ (value transformed from $E(V-I)$ using Fouqu\'e \etal 2003 coefficients). We note that decrease of the average OGLE reddening from $E(B-V)=0.143$ to $E(B-V)=0.10$ leads to decrease of absorption from $A_V=0.47$ to $A_V=0.33$ (assuming relation of Fouqu\'e \etal (2003): $A_V=3.3E(B-V)$) and hence shifts the zero point towards fainter magnitudes by $0.14$. This leads to a very good agreement with Cacciari and Clementini (2003) value. 

{\small
\begin{table}
\begin{center}
\caption{The incidence rates of Blazhko variables in the LMC. The three Tables, A, B and C, correspond to the different manners of division of stars into groups, as described in text.}

\vspace{0.4cm}

\begin{tabular}{ccccc}
\hline
 & $\mathrm{[Fe/H]}$ & $N_{BL}$ & $N_{S}$ & \%\\
\hline
A: & & & & \\
\hline
 & $\le -1.37$ & $235$ & $989$ & $19.2\pm 1.1$\%\\ 
 & $(-1.37,\ -1.19\rangle$ & $236$ & $1460$ & $13.9\pm 0.8$\% \\
 & $>-1.19$ & $231$ & $1541$ & $13.0\pm 0.8$\%\\
\hline
B: & & & & \\
\hline
 & $\le -1.32$ & $308$ & $1352$ & $18.6\pm 1.0$\%\\
 & $(-1.32,\ -1.16\rangle$ & $197$ & $1310$ & $13.1\pm 0.9$\%\\
 & $>-1.16$ & $197$ & $1328$ & $12.9\pm 0.9$\%\\
\hline
C: & & & & \\
\hline
 & $< -1.42$ & $183$ & $702$ & $20.7\pm 1.4$\%\\
 & $\langle -1.42,\ -1.02)$ & $436$ & $2644$ & $14.2\pm 0.6$\%\\
 & $\ge-1.02$ & $83$ & $644$ & $11.4\pm 1.2$\%\\
\hline
\end{tabular}
\end{center}
\end{table}
}

\section{Conclusions}

In order to examine the large data samples of the OGLE database: the Galactic Bulge and the LMC RR~Lyrae samples, we have calibrated to the {\it I}-band the photometric method of determining the [Fe/H] of the RRab star, proposed by Kov\'acs and Zsoldos (1995). Our calibrating sample consist of 28 singly periodic, galactic RRab stars. The best two-parameter formula contains the period and the Fourier phase, $\phi_{31}$, just as in case of using {\it V} light curves. However the three-parameter formula, containing also the amplitude $A_2$, was found to be a better [Fe/H] estimator, at least in the Galaxy. In case of the LMC, only the two-parameter formula gives the values of [Fe/H] which are consistent with direct spectroscopic determinations. It is a consequence of slightly different pulsational properties of the Galactic and the LMC stars. Only the two-parameter formula should be used for the LMC stars. 

Average metallicity of the Galactic Bulge RR0-S stars is $-1.04\pm 0.03$, while for the LMC it is lower, $-1.218\pm 0.004$. In both populations spread of individual metallicity values is larger than the error of the method, indicating intrinsic metallicity spread of these populations (of the order of $0.16$ for the Galactic Bulge and $0.19$ for the LMC).

We showed, that the {\it V} photometric method, may be used to obtain reliable values of Blazhko variables metallicity, if the Fourier parameters of the light curve, prewithened with frequencies of additional componenets, are used. We assume that it also holds for the {\it I}-band, used thorough this paper. Also here, only the two-parameter formula should be used, to determine the metallicity of the Blazhko variables. It is a consequence of Blazhko phenomenon dynamics, that leads to the change of amplitudes of harmonics of the fundamental mode. Average amplitudes are smaller than for the singly periodic variables, both in the Galactic Bulge and in the LMC.

In the Galactic Bulge, the incidence rate of Blazhko effect, doesn't depend on metallicity, and is equal $\sim 25-30\%$ in the whole metallicity range, $(-1.4,-0.8)$. In the LMC, Blazhko variables prefer lower metallicity values. For $\mathrm{[Fe/H]}<-1.4$ the incidence rate is significantly higher ($\sim 20\%$), while for higher metallicity values, it is approximately constant ($\sim 13\%$). Thus, the higher incidence rate of Blazhko variables in the Galactic Bulge, as compared to the LMC, cannot be attributed to the metallicity difference between the two stellar systems, as was suggested by Moskalik and Poretti (2003).

Considering the global properties of Blazhko and singly periodic variables, we note that Blazhko pulsators prefer shorter periods, with strongly decreasing incidence rate for $P>0.6$ days. Fourier phases and harmonics of the amplitudes are lower, than for the singly periodic variables, as was also noticed by Alcock \etal (2003) for their MACHO LMC sample. The relations among Fourier amplitudes are disturbed in comparison with singly periodic variables.

Estimation of metallicity for thousands of RR Lyrae stars allowed us to investigate the luminosity--metallicity relation with very high statistical precision. We found a steep dependence of $M_V$ on [Fe/H], with the slope equal $0.30\pm 0.02$. Our estimation of zero point of this reltion agrees with the literature values, if we assume that OGLE $E(B-V)$ reddenings are overestimated by $\sim 0.04$ mag.  

\Acknow{It is a pleasure to thank Pawe\l{} Moskalik for many fruitful discussions and help with improving this paper. I also thank Janusz Ka\l{}u\.zny, Beata Mazur and Kevin Lee for providing their data on the $\omega$ Cen, NGC 6362 and field variables and Alexander Schwarzenberg-Czerny for providing the Zangwill's minimisation algorithm.} 

\section{APPENDIX: Analysis of the Blazhko variables from the LMC}
We describe each Fourier series representing the Blazhko variable light curve, with the following sequence: $\bar{N}=(N_{0},\ N_{-},\ N_{+},\ N_{\Delta})$, hereafter shortened to $\bar{N}$. $N_{0}$ denotes a number of primary frequency harmonics increased by one, $N_{-/+}$ stands for the number of harmonics accompanied by additional lower or higher secondary frequency, also increased by one and $N_{\Delta}$ is equal 1 if a frequency $|\Delta f|$ is detectable, or 0 otherwise. Only terms with $A/\sigma_A>4$ are considered. In order  to find a sequence $\bar{N}$ that represents the Fourier sum, fitting the data best, the automatic procedure was developed. Conditions described below, were chosen, so that the algorithm works most effectively for the data for 30 pulsators of the Galactic Bulge sample, being strictly of BL1 or BL2 type. This test is described at the end of the Appendix. The procedure can be described in the following steps:
\begin{itemize}
\item[1] The OGLE frequencies of fundamental mode, \verb+f0+, and of additional component of the highest amplitude, \verb+f1+, are used as an input data. Frequency separation, \verb+Df=f1-f0+, is calculated.  
\item[2] In the main loop of the procedure, all sensible $\bar{N}$ are fitted to the data. Following indices goes through: \verb+n0=3,7+; \verb+np=0,n0++\verb+1+; \verb+nm=0,n0++\verb+1+; \verb+nD=0,1+; \verb+ins=0,1+. \verb+n0+ corresponds to $N_0$. If \verb+n0=7+ is found, also higher values are checked. \verb+np+ is a number of components of the form \verb+i+$\cdot$\verb+f0+$-$\verb+Df+, while \verb+nm+ is a number of components of the form \verb+i+$\cdot$\verb+f0+$+$\verb+Df+. Depending on the \verb+Df+ sign they correspond to $N_+$ or $N_-$. \verb+nD+ corresponds to $N_{\Delta}$. \verb+ins+ is equal 1 if the instrumental term of period 50,000 days is considered or 0 otherwise.  
\item[3] All the bad data points rejected in the previous steps are recovered. For each $\bar{N}$ being tested, bad data points are rejected independently.
\item[4] The multiperiodic Fourier series with the frequencies determined by $\bar{N}$, is fitted to the data. Dispersion of the fit, \verb+sig+, is calculated.
\item[5] Bad data points (threshold \verb+4sig+) are rejected using following criterions:\\
-- First bad data point is always rejected, then procedure goes to step 4.\\
-- If at least one bad data point was already rejected, the next one is rejected only if \verb+np=0+. If \verb+np+$\ne$\verb+0+ the point is kept in the sample. Please, notice that when \verb+np=0+ and \verb+nm+$\ne$\verb+0+ we check for the frequency in the location it is expected (the \verb+Df+ sign gives us information about location of the additional frequency -- on the right or left of the primary frequency). This is the way we protect the procedure, from finding BL1 type with additional componenet on the wrong side of primary frequency, by rejecting to many data points.\\
-- The above condition doesn't apply, if the correct $\bar{N}_0$ sequence was found (in the second loop of the algorithm).
\item[6] It is checked whether for all the terms, $A/\sigma_{A}>4$. If not, the procedure goes to step 2 and the next sequence $\bar{N}$ is considered. If all the terms satisfy the above condition, it is checked whether their number is greater by 2 or more, from the number of frequencies of the previously remembered sequence. If not, the sequence is remembered, otherwise not. It protects the procedure from finding apparently good solutions by fitting the Fourier series of to high order.
\item[7] From all the remembered sequences the one with the smallest \verb+sig+, $\bar{N}_0$, is chosen. The procedure is started again, but only the frequencies determined by the $\bar{N}_0$ are fitted. There is no limit on the bad data point rejection and the initial frequencies are adjusted using the Zangwill (1969) conjugate directions procedure.    
\item[8] Procedure goes to step 1 and the next star is examined.
\end{itemize} 
The algorithm was tested (without frequency adjustment) on the 30 Galactic Bulge, strictly BL1 or BL2 pulsators, described by Moskalik and Poretti (2003) by simple $\bar{N}$. As an input we used the primary frequency and one secondary frequency of the highest amplitude. Only in two cases different sequences were found. For BW10\_V44 the (4, 0, 3, 1) sequence was found, instead of (4, 0, 4, 0), but it didn't lead to significant change of Fourier parameters. The only serious problem was accounted for BW10\_V41, were the BL2 type (4, 3, 3, 0) was found, instead of a BL1, (4, 0, 3, 0), which led to the significant change in the Fourier parameters.

Based on these tests we conclude, that in vast majority of cases, our algorithm is satisfactory to obtain reliable Fourier parameters for the fundamental mode.

\newpage

\begin{figure}
\includegraphics[width=12cm]{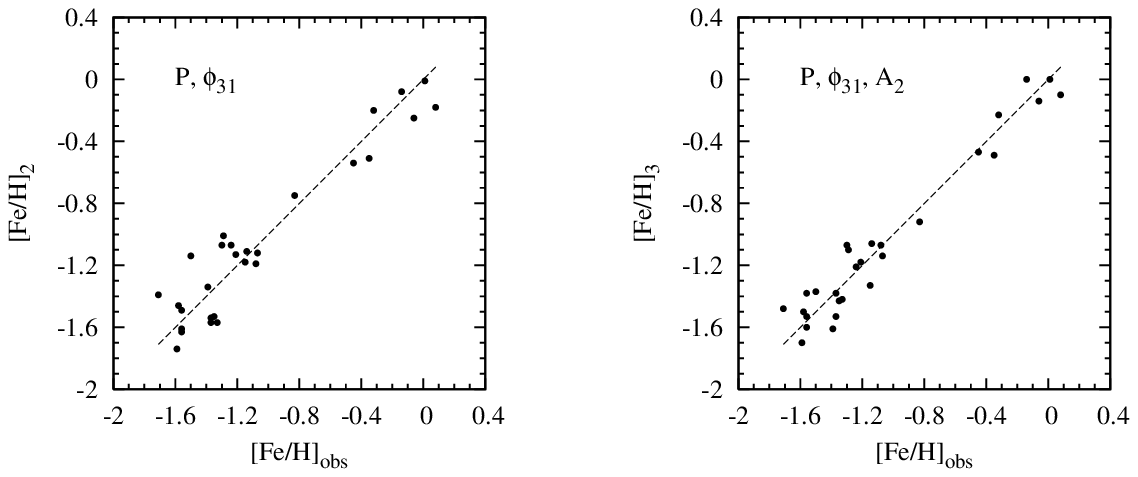}
\caption{Spectroscopic \vs photometric [Fe/H] estimated with two (left panel) and three-parameter calibration (right).}
\end{figure}

\begin{figure}
\includegraphics[width=12cm]{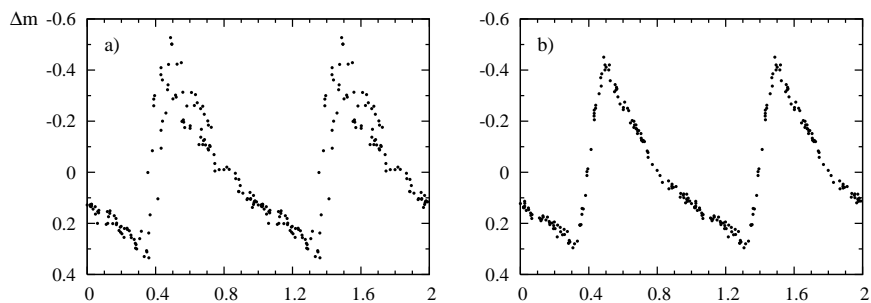}
\caption{The light curve of the RR0-BL2 pulsator, BW6\_V7 (OGLE-1 sample); a) light curve phased with the fundamental mode frequency, b) light curve after the secondary frequencies have been filtered out.}
\end{figure}

\begin{figure}
\includegraphics[width=12cm]{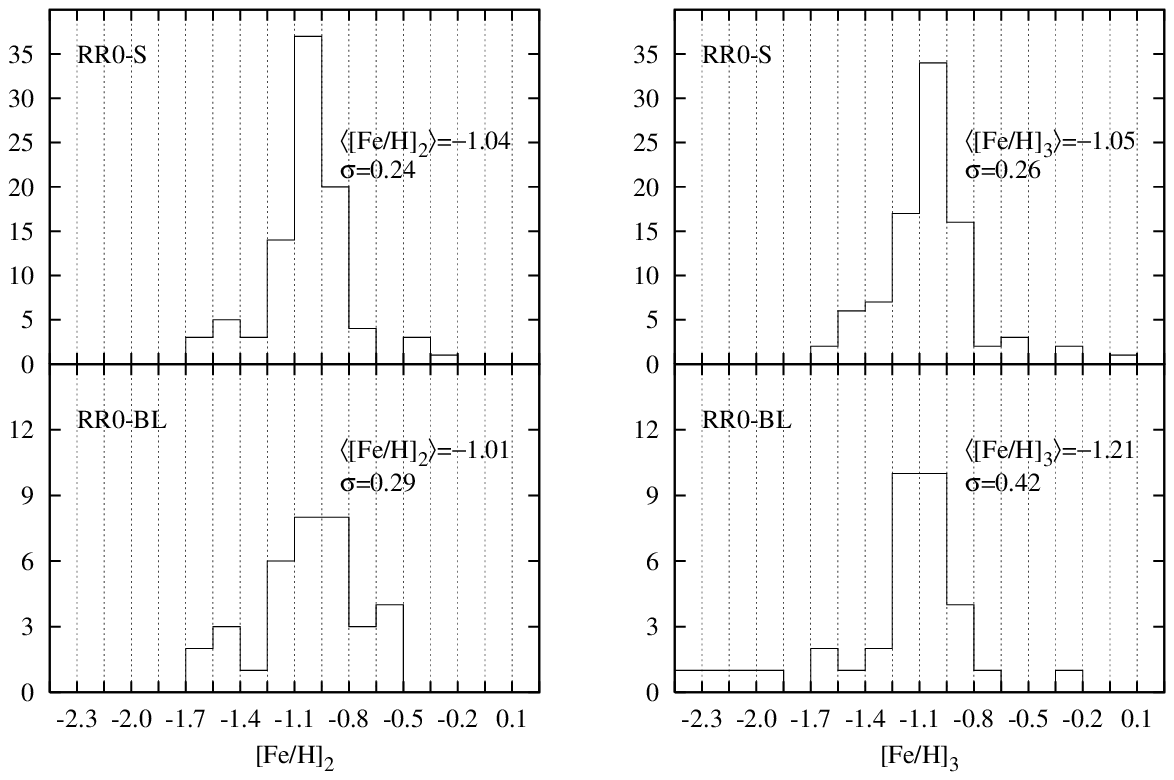}
\caption{The [Fe/H] distributions obtained with two-parameter (left) and three-parameter formula (right) for RR0-S (top) and RR0-BL (bottom) pulsators. For each distribution the average abundance, $\langle\mathrm{[Fe/H]}\rangle$, and the sample dispersion, $\sigma$, is given.}
\end{figure}

\begin{figure}
\includegraphics[width=12cm]{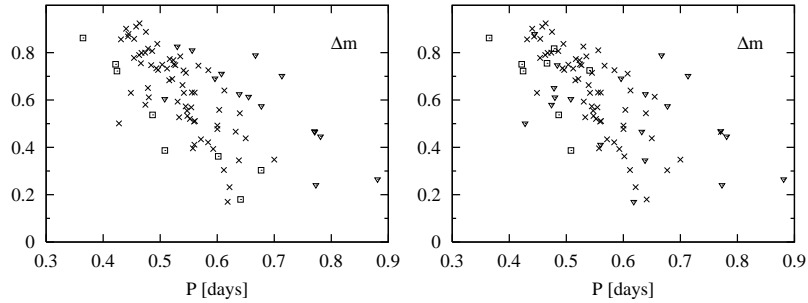}
\caption{The peak-to-peak amplitude, $\Delta m$, \vs period for the Galactic Bulge RR0-S stars of different [Fe/H], estimated with two (left panel) or three (right panel) parameter formula. Stars with different [Fe/H] are plotted with different symbols: ($\scriptstyle \boxdot$) $\mathrm{[Fe/H]}>-0.8$; ($\scriptstyle \times$) $-1.2<\mathrm{[Fe/H]}<-0.8$; ($\scriptstyle \bigtriangledown$) $\mathrm{[Fe/H]}<-1.2$.}
\end{figure}

\begin{figure}
\includegraphics[width=12cm]{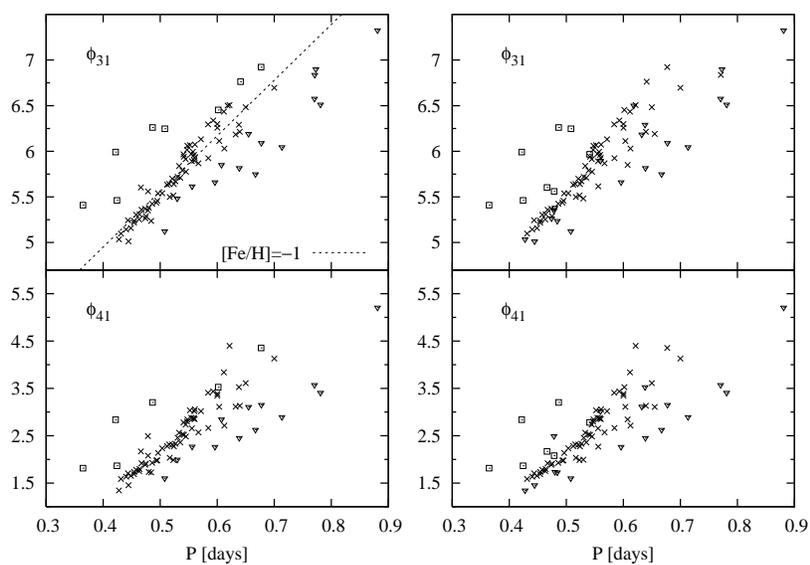}
\caption{The Fourier phases, $\phi_{31}$ and $\phi_{41}$, \vs period for the Galactic Bulge RR0-S stars. Abundances are estimated with two (left) or three-parameter formula (right). Line of $\mathrm{[Fe/H]}=-1$ is plotted. Symbols are the same as in Fig.~4}
\end{figure}

\begin{figure}
\includegraphics{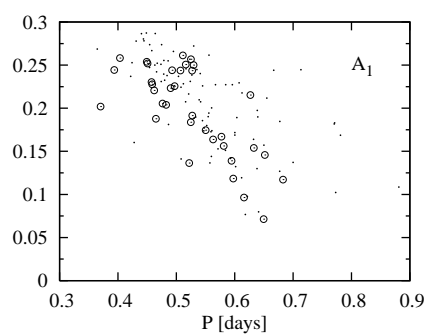}
\caption{The Fourier amplitude, $A_{1}$, \vs period for RR0-S ($\cdot$) and RR0-BL ($\scriptstyle\odot$) pulsators of the Galactic Bulge.
}
\end{figure}

\begin{figure}
\includegraphics[width=12cm]{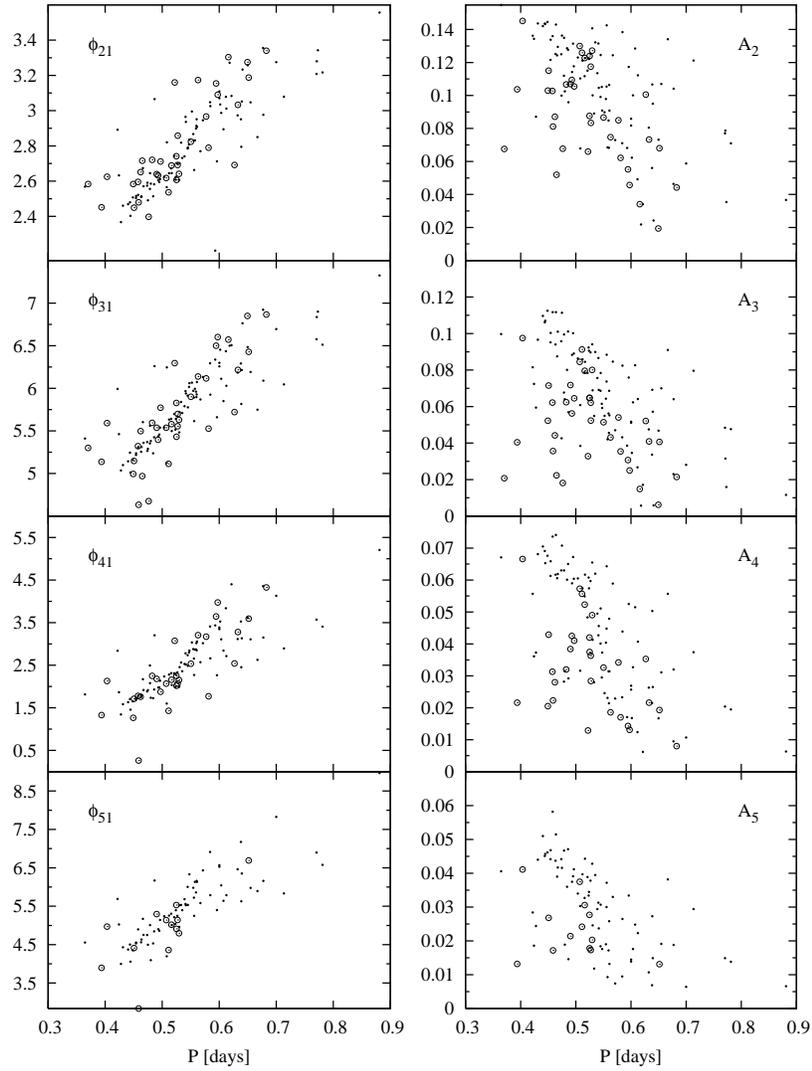}
\caption{The Fourier amplitudes and phases \vs period for the Galactic Bulge RR0-S and RR0-BL pulsators. Symbols are the same as in Fig.~6}
\end{figure}

\begin{figure}
\includegraphics[width=12cm]{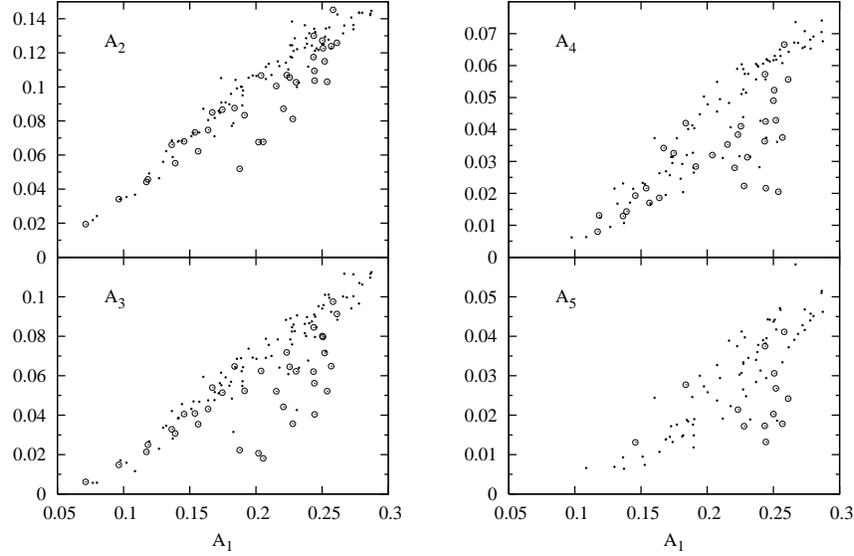}
\caption{
The Fourier amplitudes, $A_{2}-A_{5}$, \vs the Fourier amplitude of the fundamental mode, $A_{1}$, for the Galactic Bulge pulsators. Symbols are the same as in Fig.~6}
\end{figure}

\begin{figure}
\includegraphics[width=12cm]{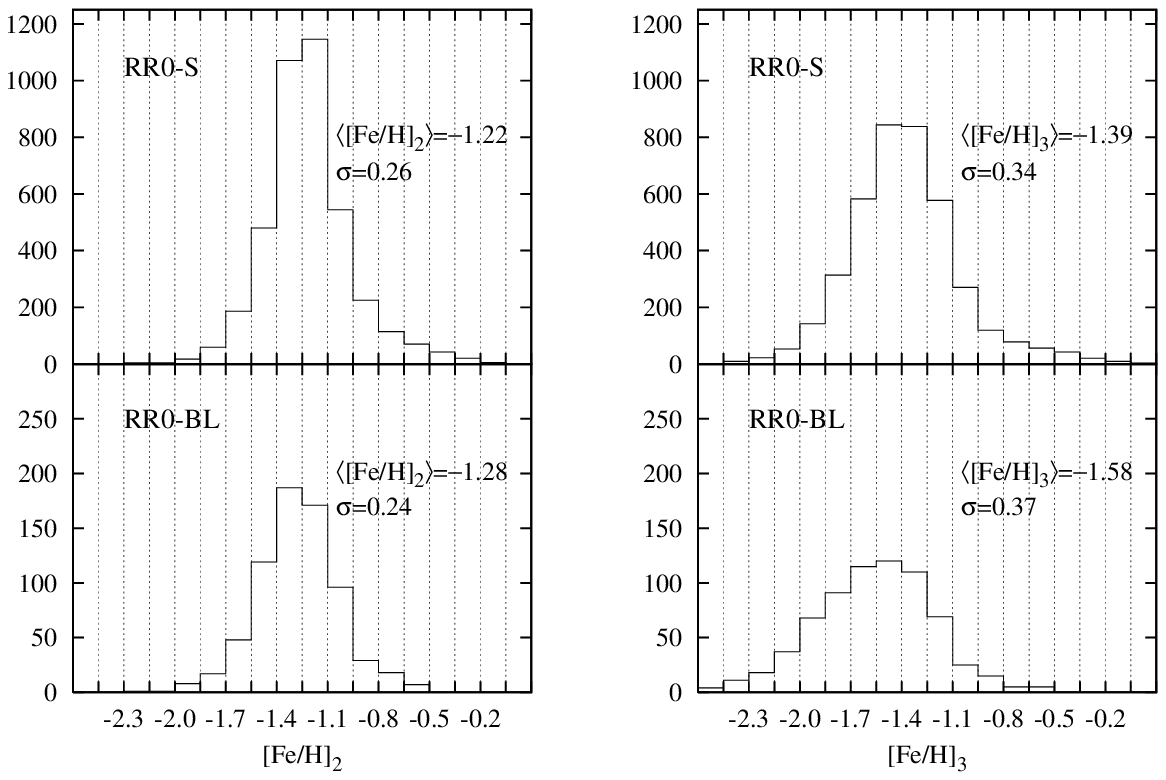}
\caption{The [Fe/H] distributions obtained with two-parameter (left) and three-parameter formula (right) for RR0-S (top) and RR0-BL (bottom) LMC pulsators. For each distribution the average abundance, $\langle\mathrm{[Fe/H]}\rangle$, and the sample dispersion, $\sigma$, is given.}
\end{figure}

\begin{figure}
\includegraphics[width=12cm]{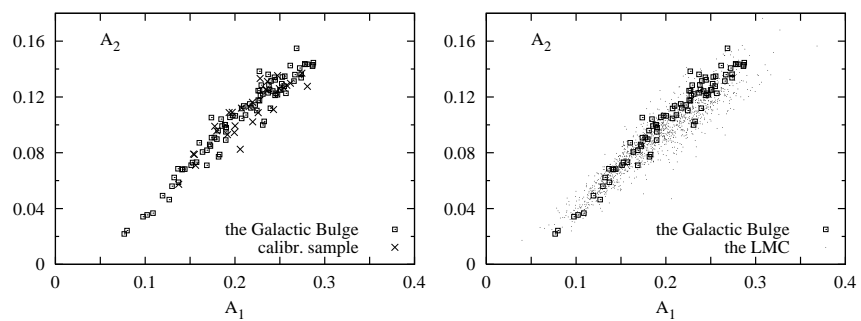}
\caption{The comparison of $A_{2}$ \vs $A_{1}$ relations for the calibrating sample, the Galactic Bulge and the LMC. For clarity one third of the LMC stars are plotted.}
\end{figure}

\begin{figure}
\includegraphics{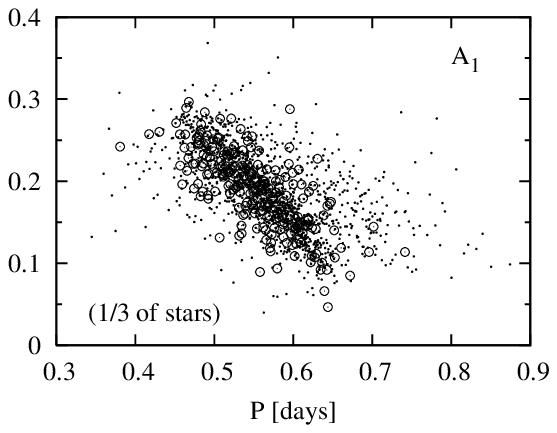}
\caption{The Fourier amplitude, $A_{1}$ \vs period, for RR0-S ($\cdot$) and RR0-BL ($\scriptstyle\odot$) pulsators of the LMC.}
\end{figure}

\begin{figure}
\includegraphics[width=12cm]{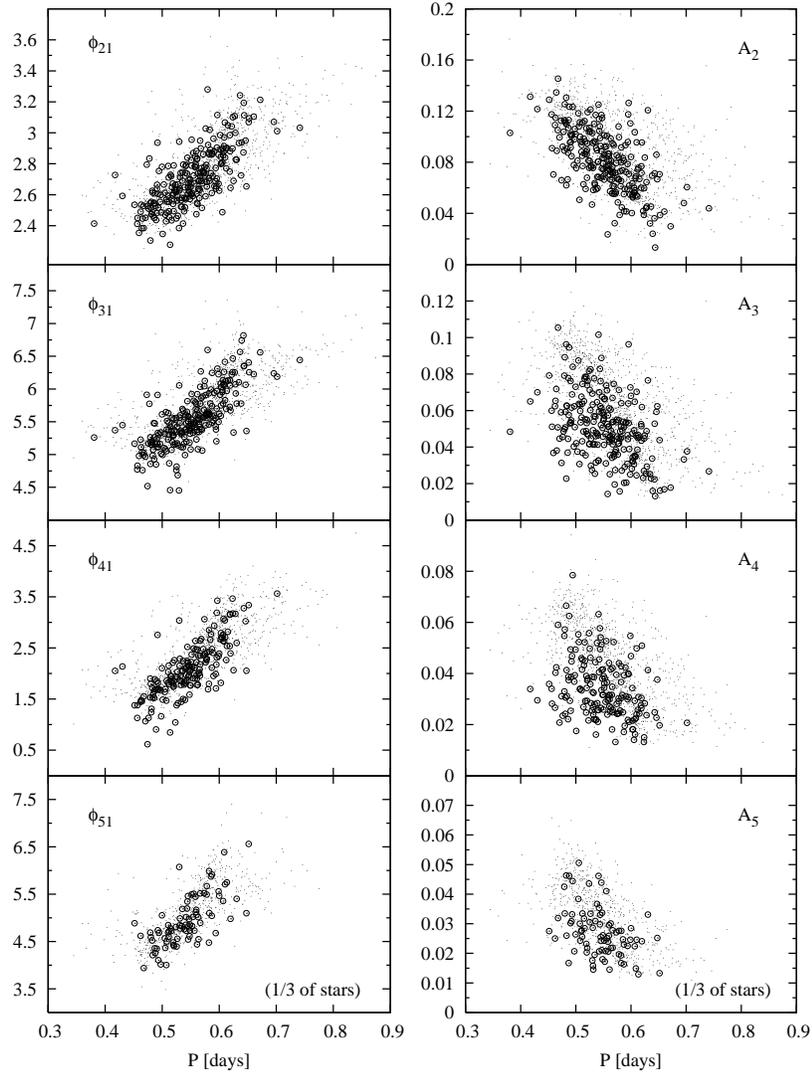}
\caption{The Fourier amplitudes and phases \vs period for the LMC RR0-S and RR0-BL pulsators. Symbols are the same as in Fig.~11}
\end{figure}

\begin{figure}
\includegraphics{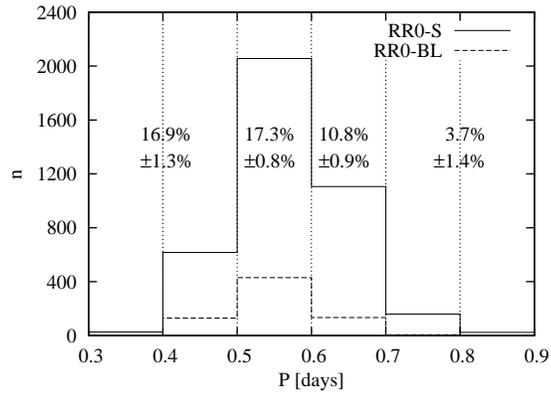}
\caption{Distribution of periods for RR0-S and RR0-BL pulsators of the LMC. The Blazhko variables incidence rate is given for four period bins.}
\end{figure}

\begin{figure}
\includegraphics{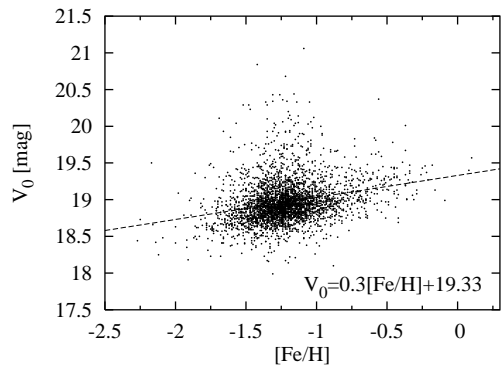}
\caption{The extinction free apparent magnitude \vs metallicity relation for the LMC RR0-S pulsators.}
\end{figure}

\end{document}